# Non destructive testing of actively cooled plasma facing components by means of thermal transient excitation and infrared imaging

R. Mitteau, S. Berrebi*, P. Chappuis, Ph. Darses*, A. Dufayet, L. Garampon, D. Guilhem, M. Lipa, V. Martin, H. Roche

Association EURATOM-CEA, Département de Recherches sur la Fusion Contrôlée

Centre de Cadarache, 13108 Saint Paul lez Durance Cedex, France

* CEDIP, 19 Bd Bidault, 77183 Croissy Beaubourg, France

SATIR is a new test-bed installed at Tore Supra to perform non destructive examination of actively cooled plasma facing components. Hot and cold water flow successively in the cooling tube of the component and the surface temperature is recorded with an infrared camera. Defects are detected by a slower temperature response above unbrazed areas. The connection between temperature differences and defect sizes is the main difficulty. It is established by tests of standard defects and thermal transient calculations of defective geometries. SATIR has been in use for two years and has proved to be very valuable to test industrial components as well as prototypes.

## 1. INTRODUCTION

The in situ maintenance of plasma facing components (PFC) is very difficult and a high level of reliability has to be reached. Therefore, non destructive examination (NDE) is systematically applied to test components manufactured by industry. Thermal techniques present the benefit to test what really matters : the thermal transfer of the bonds. Such techniques have already been in use at Tore Supra as well as in other laboratories [1,2]. At Tore Supra, the necessity for a permanent NDE test-bed for PFCs led to the development of SATIR (figure 1). SATIR is an acronym for Station d'Acquisition et de Traitement InfraRouge. This equipment and its use are described in the following pages.

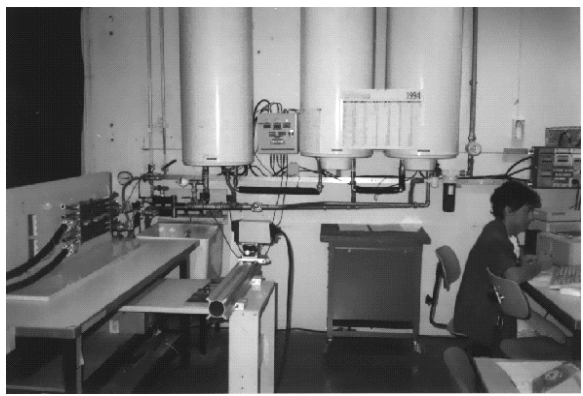

**Figure 1 : View of the test-bed**

## 2. TEST - BED DESCRIPTION

The majority of Tore Supra's PFCs are actively cooled. It gives the opportunity to use their cooling channels to heat or cool the components with hot and cold water. An open water circuit is set up (figure 2). Cold water comes from the commercial network and three heating tanks totalling 600 l are installed to deliver hot water. Two elements can be installed in parallel. Various types of connections can be used. This thermal excitation is highly efficient. For a temperature difference of 80 °C between the component and the water, and with an heat transfer coefficient to 17000 W/m²K, the wall heat flux reaches 1.4 MW/m².

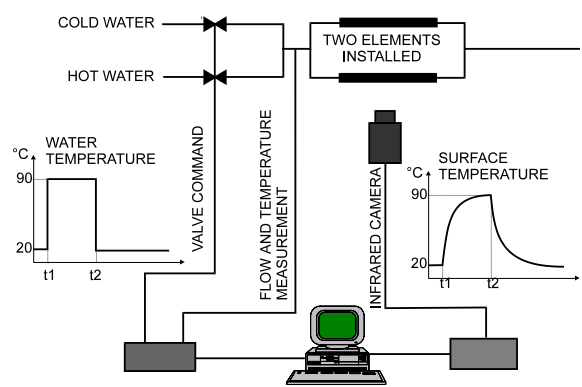

**Figure 2 : Test-bed diagram**

Depending on the pressure drop of the elements, the hot water flow rate amounts up to 2.5 m³/h and the cold water flow rate up to 4.3 m³/h. These flow rates are sufficient to have a water transit duration through the element (~ 0.1 s) shorter than the thermal time constant of the element (3 to 15 s).

The surface temperature of the elements is measured with an infrared camera (type Inframetrix 600). The video signal is numerised and stored in a PC. Safety copy of the film may be stored with a VCR. The PC works with a PTR-based software (CEDIP) which both remote controls the acquisition sequence as well as it does the thermal analysis. The PC has a 486 processor and 64 Mo RAM which enable to record up to 12.5 images per second. The data are safeguarded on 5 Go cartridges.

## 3. TEST PROCEDURE

The choice of hot and cold water durations and sampling frequency depends on the element being tested. The inner first wall whose thermal time constant is 15 seconds was tested with 60 seconds fronts and a sampling frequency of an image out of 6 (roughly 2 images per second) [3].

Mock-up aimed at developments are tested individually and get a customised analysis. Elements from large fabrication series are tested simultaneously with a sound reference element, which is chosen after test in a high heat flux test-bed. The subsequent analysis relies on the comparison between the two.

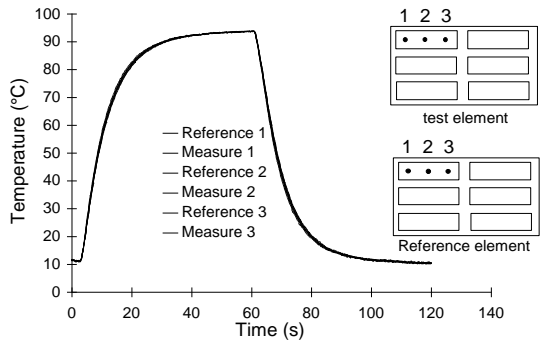

**Figure 3 : Surface temperature of 6 points : the 6 lines are merged**

The thermal analysis is based on the difference between the time response of each couple of points (test and reference, figure 3). These points are located at the same relative positions on the tiles to allow comparison. The temperatures are extracted from the film, usually from a 3*3 pixel matrix. Figure 3 does not permit to distinguish differences in the surface temperature. One has to display the differences to see the differences (figure 4.a).

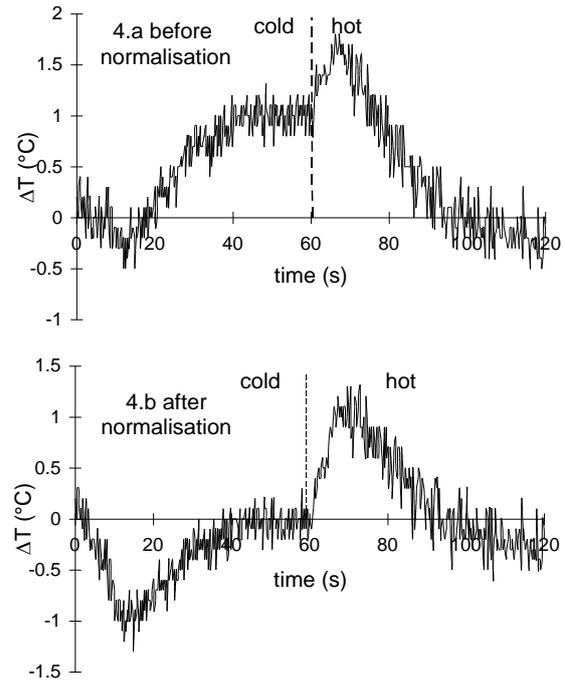

**Figure 4 : Temperature difference with and without normalisation**

Because of emissivity differences at the surface of the tiles, stable hot and cold temperatures measured by the camera may vary from point to point. In order to calculate the real temperature difference between the points, the temperature curves are normalised linearly (figure 4.b).

Let $T_c$ and $T_h$ be the cold and hot stable temperature as measured by the infrared camera, $\tilde{T}_c$ and $\tilde{T}_h$ the cold and hot stable temperatures averaged on all curves. The normalised temperature is given by :

$$T_{normalized} = (T_{measured} - T_c) \cdot \frac{(\tilde{T}_h - \tilde{T}_c)}{(T_h - T_c)} + \tilde{T}_c$$

Maximum temperature differences are then deduced from the normalised curves and displayed on the screen (see example table 1). In the case of the inner first wall, the temperature differences were also corrected from the various wall thicknesses measured on the stainless steel heat sink.

A maximal temperature difference is authorized. When an element shows a temperature lag that exceeds the limit, it is more thoroughly investigated and can be rejected.

| | ΔT up | ΔT down | defect size |
|---|---|---|---|
| pair N°1 | 1.3 °C | 1.3 °C | 2.8 mm |
| pair N°2 | 1.1 °C | 1.6 °C | 3.2 mm |
| pair N°3 | 1.3 °C | 1.5 °C | 3.2 mm |

**Table 1 : Results of the thermal analysis**

Setting the limit is the greatest difficulty. Two methods are employed : test of standard defects and finite element calculations.

The standard defects are either fabricated (e.g. during the brazing cycle by forbidding the braze to wet the armour material using stop-off fluid, figure 5) or created on sound elements (by drilling or grinding the joint with narrow tools). The elements are then tested and the temperature differences plotted against defect sizes.

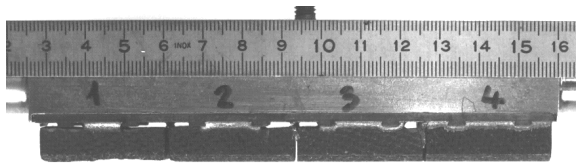

**Figure 5 : Standard defects**

Finite elements calculations of faulty geometries are also extensively performed. Both 2D and 3D calculations are made. They are compared to the results of the standard defects. The calculations showed that the temperature difference on the surface of the tile is better correlated to the braze void extension rather than to the braze void area. When no boundary is present, the braze void extension is the radius of the largest circle that can be inscripted in the defect (figure 6).

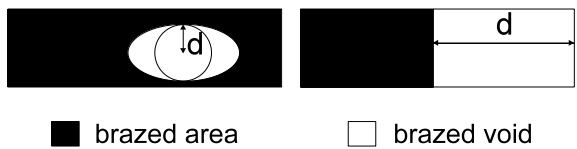

**Figure 6 : illustration of braze void extension**

However, with boundaries, a more complicated definition has to be used. If we consider the shortest path between a point of the braze void and all points of the brazed area, the braze void extension is the longest of the those paths. This can be mathematically written by the following expression :

$$d = \max_{M \in (\text{braze void})} \left( \min_{N \in (\text{brazed area})} (MN) \right)$$

Overheating of the tile's surface under heat flux is governed by the same parameter, so that setting an acceptance limit on the test bed is equivalent to accept a limited overheating under heat flux. For the inner first wall, those considerations led to a limit of 6°C ([3], figure5).

## 4. TWO YEARS EXPERIENCE

The experience was gained mainly on the inner first wall, which elements were tested after delivery and after the assembly steps. Testing 1 m² takes approximately 1 month. The test-bed led to the rejection of three elements. One tile had an 11°C lag (figure 7).

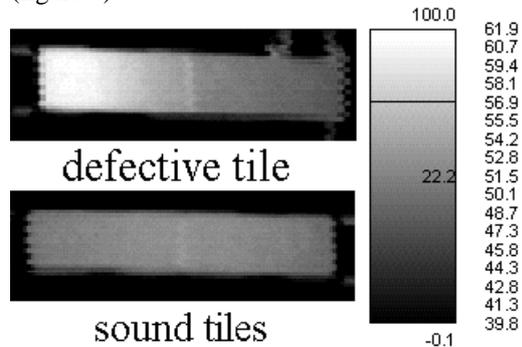

**Figure 7 : Defect on a PPI element**

X-ray testing of this element had shown no defect, but cuts through the element confirmed the presence of a large defect. This example proved the usefulness of SATIR and its complementarity to X-ray examination.

Many mock-up were also individually tested (fingers, macroblocs, bolted tiles, metallic mock-ups, see table 2). They showed that, as in other NDE techniques, experience is vital and allows to have a better sensibility.

## 5. FUTURE PROSPECTS

The software is currently being improved to analyse highest and lowest temperatures on surfaces rather than on points. The aim is to detect surface temperature's differences within single tiles. By so doing, the analysis will be independent from temperature lag caused by conductivity or thickness differences between the measured and reference tiles. This will allow to reduce the acceptance limit close to the level of the signal's noise. A comparison to a reference element will however be maintained, to avoid the risk of generalized defects. These techniques will be used to test the high heat flux fingers that are developed for the Tore Supra's Toroidal pump limiter. A 3°C limit is foreseen, which would guaranty a steady state temperature smaller than 1500 °C (in the worst case, figure 8). This value is correlated to a 6 mm defect.

| component | Nb | Nb tiles | area (cm²) | tests | results | remarks |
|---|---|---|---|---|---|---|
| Tore Supra inner first wall (IFW) | 88 | 1506 | 19500 | 2/3 | 3 elements rejected | 60° toroïdal extension 6 months |
| Tore Supra IFW standard defects | 1 | 4 | 24 | 1 | ΔT function of void percentage | correlated to FE calculations |
| LPT short fingers | 8 | 56 | 350 | 2 | defaults hardly visible before FE200, easily visible after damage on FE200 | studies are under progress to improve the analysis and lower the acceptance limit |
| LPT long fingers | 4 | 84 | 500 | 2 | idem | idem |
| HIP copper - Stainless Steel (ITER) | 1 | no tile | 100 | 1 | non homogenous copper emissivity | task T8 |
| rheocast (ITER) | 1 | 1 copper tile | 30 | 1 | very large defect | deteriorated during tests at FE200 |
| macrobloc monotube | 3 | 1 large | 30 | 1 | evidence of braze voids | each face has to be tested separately |
| macrobloc multitube | 2 | 1 large | 300 | 3 | | |
| macrobloc | 2 | 1 large | 300 | 1 | to be tested | (recessed hole) |
| ergodic divertor neutraliser | 1 | $B_4C$ coating | 100 | 3 | the thickness of the $B_4C$ coating prevails | no thermographic test for this fabrication |
| ergodic divertor front face | 1 | 32 bolted CFC | 430 | 1 | the loosest tiles lag | semi-inertial element time constant 300 s |

**Table 2 : Experience gained during two years of operation**

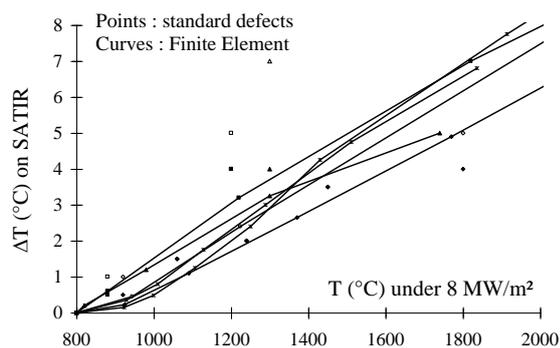

**Figure 8 : Setting the limit for the fingers**

Beside this improvement, other techniques could be used to increase both the sensibility of the test-bed and its capacity :
- Going towards higher water pressure, velocity and temperature. However, this would require a stronger water loop and stronger connections.
- Cycling the excitation and measure the phase shift.
- Narrowing the temperature range of the camera to increase the precision of the measure. Correcting the emissivity differences on the surface of the tile would require to store a map of the surface's emissivity.

## 6. CONCLUSION

SATIR has proved to be a very valuable test - bed to test fabrication series as well as prototypes. In comparison to other thermal techniques used in NDE, SATIR presents the advantage of being quantitative. It helps judgement when elements are defective but might be accepted.